\documentclass[prl,aps,twocolumn,groupedaddress,superscriptaddress,floatfix,showpacs]{revtex4}
\usepackage{epsfig}
\usepackage{mathrsfs}
\usepackage{amsmath}
\usepackage{textcmds}
\usepackage{amssymb}
\usepackage{mathptmx}
\allowdisplaybreaks[1]
\newcommand{\be}{\begin{equation}}
\newcommand{\ee}{\end{equation}}
\newcommand{\bea}{\begin{eqnarray}}
\newcommand{\eea}{\end{eqnarray}}
\newcommand{\ket}[1]{\left|#1\right\rangle}
\newcommand{\bra}[1]{\left\langle #1\right|}

\newcommand{\expect}[3]{\left\langle #1 \right| #2 \left| #3 \right\rangle}

\newcommand{\bc}{\begin{center}}
\newcommand{\ec}{\end{center}}

\newcommand{\forget}[1]{}

\newcommand{\re}{{\rm e}}

\newcommand{\ri}{{\rm i\,}}
\begin{document}
\title{Multipartite Entanglement in Non-Equilibrium Quantum Phase Transition in a Collective Atomic System}
\author{Kishore T. Kapale}
\email{KT-Kapale@wiu.edu}
\affiliation{Department of Physics, Western Illinois University, Macomb, IL 61455}
\affiliation{Hearne Institute for Theoretical Physics, Department of Physics \& Astronomy,
Louisiana State University,
Baton Rouge, Louisiana 70803-4001}
\author{Girish S. Agarwal}
\email{Girish.Agarwal@okstate.edu}
\affiliation{Department of Physics, Oklahoma State University, Stillwater, Oklahoma, 74078}

\begin{abstract}
We study multipartite entanglement in non-equilibrium quantum phase transition (NEQPT) attainable in a coherently driven atomic ensemble undergoing collective decay. The NEQPT arises in the steady state of the system as  the drive field strength is varied in comparison with the decay parameter.   A close connection is observed between the critical behavior and the multi-partite entanglement in the system determined via the von Neumann entropy. The derivative of the von Neumann entropy shows $\lambda$-type behavior typical of the second order phase transitions. We further show that the calculated bi- and tri- partite entropies satisfy the Lieb inequality.
\end{abstract}
\pacs{03.65.Ud, 03.65.Yz, 64.60.Ht}
\maketitle 
Quantum phase transitions (QPT)---phase transitions occurring in quantum matter at zero temperature---signify critical behavior in the ground state of the system under the influence of an external physical parameter~\cite{QPT} and are difficult to study via direct methods due to large number of degrees of freedom involved. Recently, there has been a strong interest in the simulation of such complex systems (e.g., condensed-matter systems)  in simple controllable quantum systems that are simpler to understand, calculate or observe such as trapped-ions\forget{~\cite{Wineland:1998,Leibfried:2002,Schaetz:2004}}~\cite{IonTrapQS}. Some notable examples of simulation of familiar QPTs of condensed-matter systems include the observation of the superfluid to Mott insulator transition in Bose-Einstein Condensates~\cite{Greiner:2002} and study of the BCS to BEC crossover in degenerate Fermi systems~\cite{Stewart:2006}. In this context, study of QPTs, for example as observed in spin chains~\cite{Wang:2002}, which could be simulated in Dicke type systems\forget{~\cite{Schneider:2002,Lambert:2004,Lee:2004,Reslen:2005}}~\cite{DickePT} or optical lattices~\cite{Garcia-Ripoll:2004}, is especially important to understand the connection between entanglement and the critical behavior. It has also been noted that the universality of the critical phenomena simply extends to the pairwise-entanglement~\cite{Osterloh:2002,Korepin:2004}. Further, there have been extensive studies of the so-called Lipkin model involving chain of fermions with short-range interactions\forget{~\cite{Garg:1993,Dusuel:2004,Unanyan:2005,Unanyan:2005a}}~\cite{LipkinModel}; however, optical realizations of this model are far from clear.  Study of the spin chains is further important because of possible applications in quantum communication~\cite{Bose:2003}.  It could, however, be noted that so far there has been no study relating the multipartite entanglement and critical behavior in non-equilibrium QPT (NEQPT).

Here we show how a non-equilibrium quantum phase transition      (NEQPT) can be achieved in a collective atomic ensemble and  study the relation of multiparticle entanglement with this type of critical behavior. The dynamics of the atomic system can be controlled by the external drive field, variation of which over a wide-range, in comparison with the collective decay parameter, gives rise to a phase transition in the steady state of the system. The system consists of a collectively driven ensemble of two (or three) level atoms that is simultaneously subjected to a collective emission of radiation as for example induced by a cavity. Thus, due to the continuous driving and collective decay the system is inherently a non-equilibrium system. The non-equilibrium steady state (NESS) density matrix of the system can be determined and allows  study of the entanglement of the complete system or parts of it in terms of the von Neumann entropy which has been recognized as a measure of entanglement and quantum correlations for optical systems for quite some time~\cite{Huang:1994,Barnett:1989}.

\forget{The master equation for a two level atom (see Fig. 1 (A)) undergoing radiative decay at the rate $2\Gamma$ via its interaction with a bath of vacuum modes is given by\cite{QO}
\begin{equation}
\dot{\rho} = - \Gamma (s^+ s^- \rho - 2 s^+ \rho s^- + \rho s^+ s^-)\,,
\label{Eq:SingleDecay}
\end{equation} 
where $s^+=\ket{a}\bra{b}$ and $s^-=\ket{b}\bra{a}$ are the atomic operators signifying atomic transitions. The collective decay dynamics, where all the atoms in an ensemble cooperatively decay~\cite{Dicke:1954}, can be described by replacing the individual atomic operators $s^\pm$  by their collective versions $S^\pm=\sum_i s_i^\pm$ in Eq. \eqref{Eq:SingleDecay}.}

\begin{figure}
\centerline{\includegraphics[width=0.6\columnwidth]{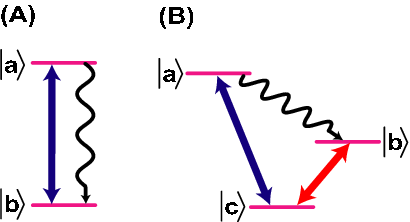}}
\caption{\label{Fig:Model} (A) Level scheme for two-level atoms: A coherent drive and collective decay into the cavity mode allow realization of the dynamics in Eq.~\eqref{Eq:DenMatDynamics}. (B) Level scheme for three-level atoms:  A coherent drive and microwave field on the lower transition alongwith collective decay into quantum vacuum of a cavity allow for a different realization of the dynamics in Eq.~\eqref{Eq:DenMatDynamics} that has been also useful for quantum communication~\cite{Duan}. In the notation of this figure $s_{i}^+ = \ket{a}\bra{b}$. }
\end{figure}

We consider a collection of two level atoms, described by the spin-half operators ($S^\pm_j, S^z_j$) and the transition frequency $\omega$, interacting resonantly with a single mode of the cavity field, described by boson operators $a$ and $a^\dagger$. The atoms are in addition driven by a coherent laser light resonant with the atomic transition. The Rabi frequency of the interaction of the atom with the coherent light is $|\Omega| \re^{\ri \phi}$ and with the cavity mode is $g$.  The total Hamiltonian is given by 
\begin{align}
{\mathcal H} =& \hbar \omega \sum_j S^z_j + \hbar \omega a^\dagger a
-\hbar \sum_j (g_j \re^{\ri \phi_j} S^+_j a + \mbox{H. c.}) \nonumber \\
&-\hbar |\Omega|\sum_j(\re^{\ri \phi} \re^{-\ri \omega t} S^+_j \re^{\ri \phi_j}  + \mbox{H. c.} )
\end{align}
The dynamics described by the above Hamiltonian is to be supplemented by the dissipative terms describing the leakage from mirrors. The cavity field decay rate is taken to be $\kappa$. In the bad cavity limit, $\kappa \gg g$, cavity field can be adiabatically eliminated to arrive at the following master equation for the atomic density matrix:
\begin{equation}
\dot{\rho} =\ri |\Omega| [ \re^{-\ri \phi} S^+ + \re^{\ri \phi} S^-, \rho]
  - \Gamma (S^+ S^- \rho - 2 S^+ \rho S^- + \rho S^+ S^-)\,.
\label{Eq:DenMatDynamics}
\end{equation}
Here $\Gamma=|g|^2/\kappa$ is the collective decay parameter, and $S^\pm = \sum_j S^\pm_j \re^{\ri \phi_j}$ are the collective atomic operators. Here the density matrix is written in a frame rotating with the frequency of the applied field $\omega$ therefore it does not appear explicitly in the dynamical equation. We add that another very popular system, three-level system with single-photon emission on the $\ket{a}-\ket{b}$ transition, which has been found to be useful for long-distance quantum communication~\cite{Duan}, could also be described by the same dynamical equation~\eqref{Eq:DenMatDynamics} (See Fig.~\ref{Fig:Model} (B)). 

The collective atomic operators are analogous to the ladder  operators for a spin $N/2$ system consisting of total $N$ two-level atoms contained in the ensemble.  It is important to note at this point that there are two competing dynamical processes going on in the system arising due to the external coherent drive and the collective decay. As it will be shown later this competition gives rise to the phase transition in the system as the relative amplitudes of $|\Omega |$ and $\Gamma$ are varied. We have recently showed how to generate generalized Werner states, which are analogues of the two-particle maximally-entangled mixed states, by coherently driving the collectively decaying ensemble of two-level atoms~\cite{Agarwal:2006}.

Despite the non-equilibrium nature of the system under consideration there exists a steady state that can be formally written as~\cite{Puri:1980} 
\begin{equation}
\rho =(R^-)^{-1} (R^+)^{-1}/D = (S^- + \ri G)^{-1}(S^+ - \ri G)^{-1}/D \,,
\end{equation} 
where  $R^- =  S^- + \ri (|\Omega|/\Gamma) \re^{\ri \phi}$,  $D=\mbox{Tr}\,[(S^- + \ri G)^{-1}(S^+ - \ri G)^{-1}]$ is the normalization and the newly introduced parameter $G$ is given by $G= \ri \Omega/\Gamma$. This allows us to determine the steady state of the system, nevertheless, it is important to know the typical time required to attain this steady state in order to determine the rate at which the system can be swept through the phase transition. In general, the time required for such a system of $N$ atoms to arrive at the steady state is roughly $\Gamma t \approx 1/N$ for any arbitrary value of $\Omega$ (See Ref.~\cite{Agarwal:2006} for discussion of a two atom case).   Thus, the time-rate of the variation of the drive field Rabi frequency, required to scan through the phase transition, needs to be smaller than the time-rate of approach to the steady state. Within this regime $\Omega$  in Eq.~\eqref{Eq:DenMatDynamics} is taken to be time-independent. The collective decay rate increases as the size of the system increases allowing faster variation of the amplitude $|\Omega|$ to study the passage of the system through the phase transition.

In the spin representation of the collective system, the steady state density matrix can be written as:
\begin{align}
\rho &= \frac{1}{D} \sum_{m,n=0}^{N} (-\ri G)^{-m} (\ri G^*)^{-n} (S^-)^m 
       (S^+)^n 
       \label{Eq:RhoSteadyState1} \\
       &=\sum_{p,q=-N/2}^{N/2} \rho_{p,q} \ket{\frac{N}{2},p}\bra{\frac{N}{2},q}\,.
\label{Eq:RhoSteadyState2}
\end{align}
Representing the spin operators in Eq.~\eqref{Eq:RhoSteadyState1} 
in the basis of angular momentum states with $S=N/2$ (so called Dicke states~\cite{Dicke:1954}) the individual density matrix elements $\rho_{p,q}$ in Eq.~\eqref{Eq:RhoSteadyState2} can be determined. This is possible as the dynamics in Eq.~\eqref{Eq:DenMatDynamics}   is invariant under the exchange of the particle labels and the state remains completely symmetric and is confined to the manifold with $S=N/2$ provided the initial state of the atoms is completely symmetric. These symmetry considerations are also important in the extraction of density matrix of the different-sized parts of the system for further study of the entanglement.

Once the density matrix for the complete system is known, the density matrix
$\rho^{(M)}$  for part of the system containing M particles can be projected out via $\rho^{(M)} = \sum_{\mu,\nu=-M/2}^{M/2}(\mbox{Tr}_{N-M} \expect{M/2,\mu}{\rho}{M/2,\nu}) \ket{M/2,\mu}\bra{M/2,\nu}$. The projected density matrix can then be used to study properties of a part of the system.  
To maintain the symmetry of the parts of the system, the state of the $N$ particle state has to be expressed as a superposition of completely symmetric parts with $M$ and $N-M$ particles. This is possible through the spin representation that we have already introduced; thus, 
$\ket{{N}/{2},p} = \sum_{p = j + k} C^{N/2}_{j,k} \ket{M/2, j} \ket{(N-M)/2, k}\,,$
where the $ C^{N/2}_{j,k}$ are the Clebsch-Gordan coefficients arising due to the coupling of the component angular momenta to form the total spin $N/2$ system~\cite{Sobelman:1996}. 
Thus, the density matrix elements $\rho^{(M)}_{\mu,\nu}= \expect{\frac{M}{2},\mu}{\rho^{(M)}}{\frac{M}{2},\nu}$ are given by
\begin{equation}
\rho^{(M)}_{\mu,\nu}= \sum_{p,q=-N/2}^{N/2} \rho_{p,q}  {\sum_{l=-(N-M)/2}^{(N-M)/2}\,\,\sum_{p = \mu+l}\,\,\sum_{q=\nu+l}C^{N/2}_{\mu,l} {C^{N/2}_{\nu
,l}}^*}\,.
\end{equation}
allowing determination of multipartite entanglement in the system.
In particular, we study bi- and tri-partite entanglement in an atomic ensemble containing total $N$ particles.

\begin{figure}
\centerline{\includegraphics[width=0.6\columnwidth]{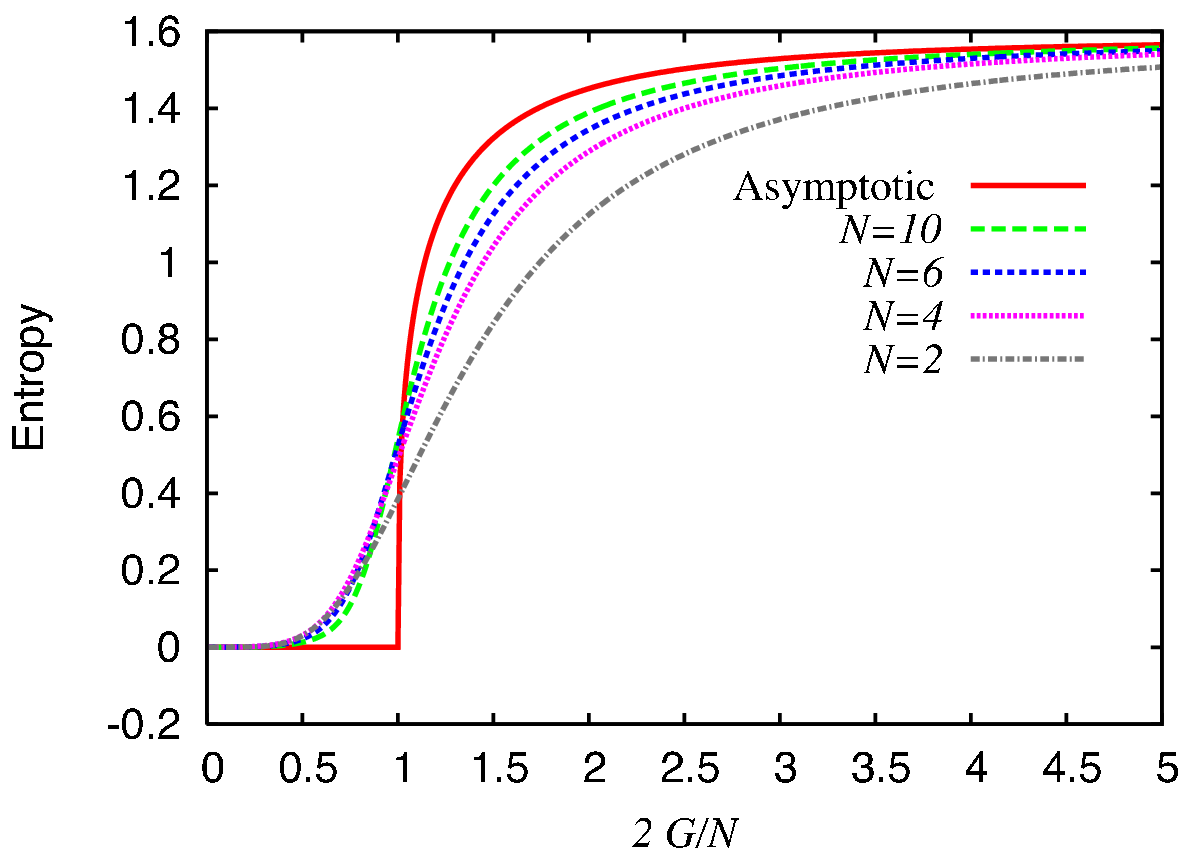}}
\caption{\label{Fig:EntropyTwoOf} Bipartite entanglement in the non-equilibrium QPT examined via the von Neumann Entropy defined in Eq.~\eqref{Eq:VNE}.}
\end{figure}

\begin{figure}
\centerline{\includegraphics[width=0.6\columnwidth]{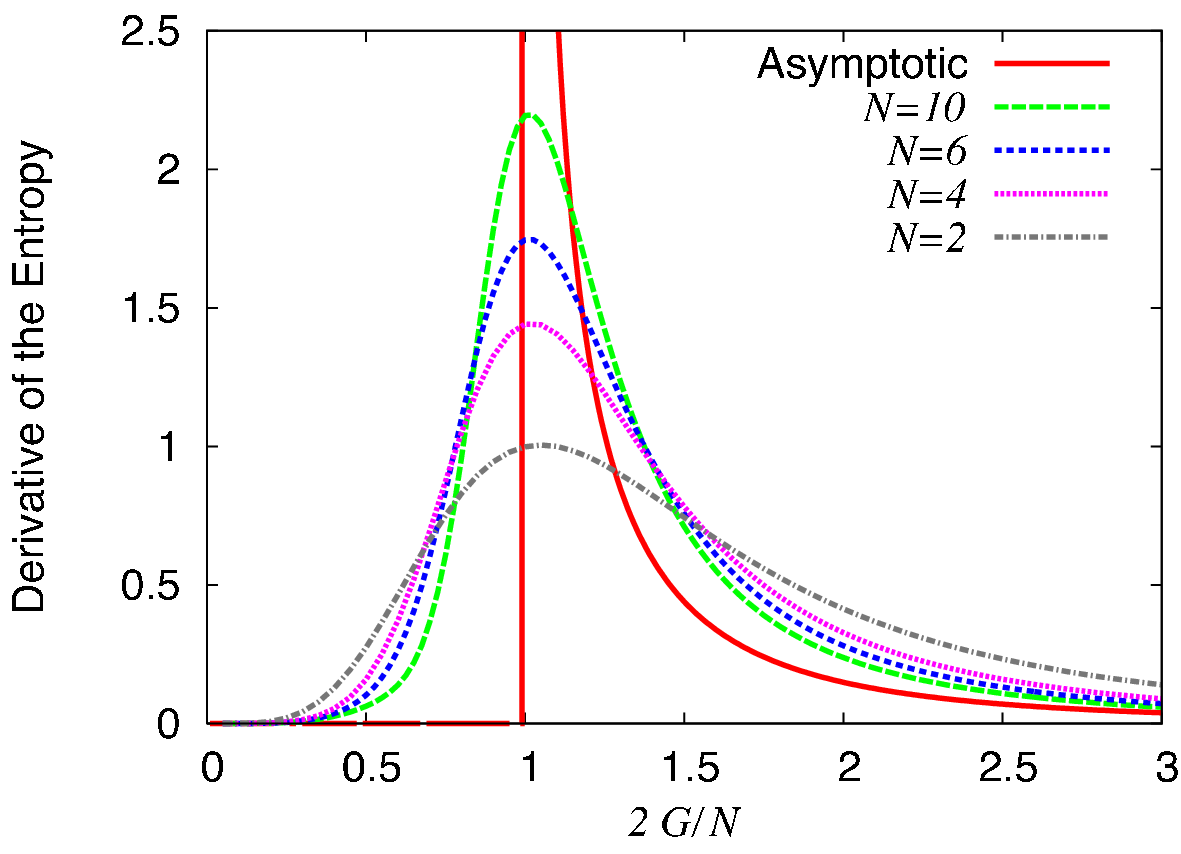}}
\caption{\label{Fig:DerivativeEntropyTwoOf} Bipartite entanglement in the non-equilibrium QPT examined via the derivative of the von Neumann Entopy. The behavior in the asymptotic limit is similar to the $\lambda$-type variation of the specific heat typical of second order phase transitions.}
\end{figure}

\begin{figure}
\centerline{\includegraphics[width=0.6\columnwidth]{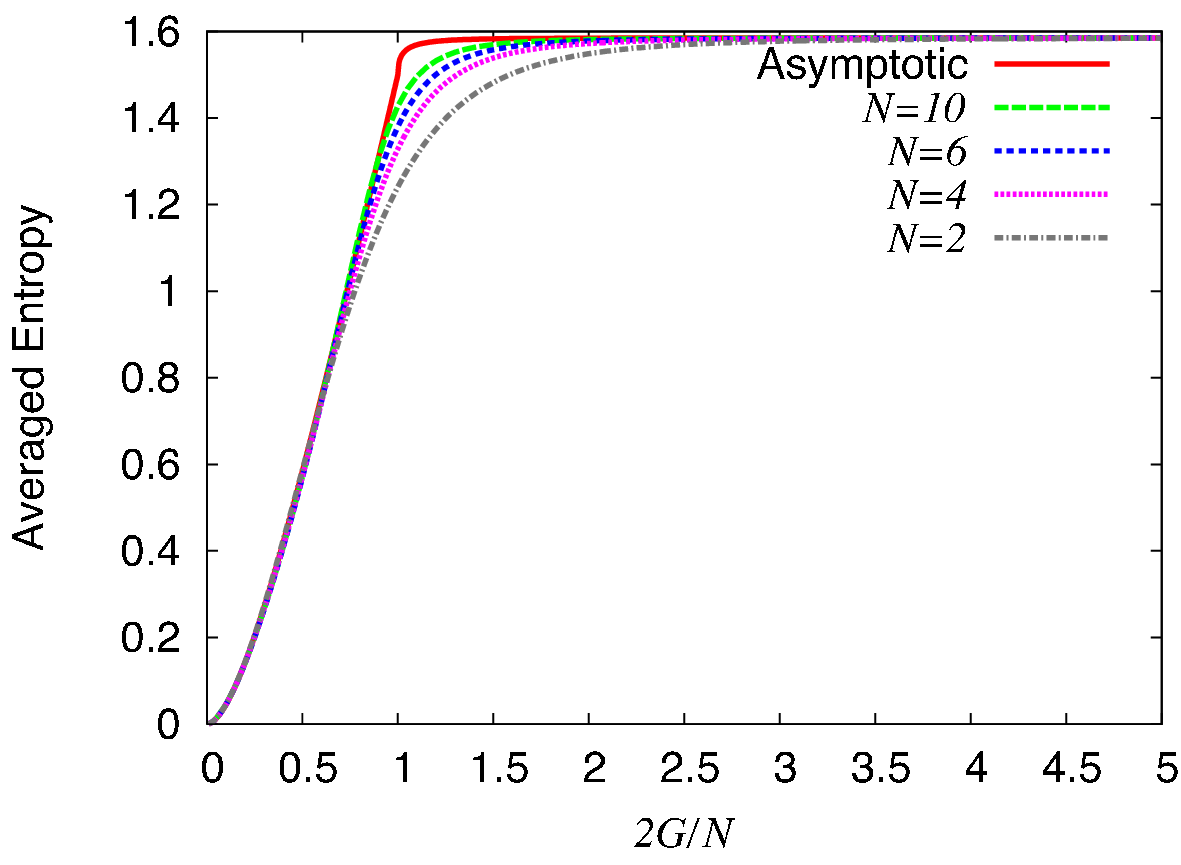}}
\caption{\label{Fig:AveragedEntropyTwoOf} Bipartite entanglement in the non-equilibrium QPT examined via the average of  the von Neumann Entropy over random phases of the driving field. There is no sharp transition even in the asymptotic limit. \forget{The visible kink is an artifact arising due to the manner in which the asymptotic limit is calculated in two parameter regimes on both sides of $2 G/N=1$.}}
\end{figure}

In  Figs.~\ref{Fig:EntropyTwoOf}-\ref{Fig:AveragedEntropyTwoOf} we show a connection between bipartite entanglement  and the critical behavior in the NEQPT observable in the collectively driven system as a function of the scaled Rabi frequency associated with the drive field $
\Omega_s = 2 G / N = 2 \ri \Omega/ N \Gamma\,.
$
The plot in Fig.~\ref{Fig:EntropyTwoOf} shows the von Neumann Entropy of the entangled state determined from the projected two-particle density matrix $\rho^{(2)}$ as~\cite{Nielsen:2000}
\begin{equation}
S(\rho^{(2)}) = - \mbox{Tr}\,(\rho^{(2)} \log_2 \rho^{(2)})\,.
\label{Eq:VNE}
\end{equation} 
It is known that the non-equilibrium phase transition occurs in these systems at $\Omega_s = 2 G/N =1$~\cite{Puri:1980,Narducci:1978}; we observe that the entanglement character of the two-partite entanglement in the $N$ particle system follows the phase transition. In the asymptotic limit of $N, G \rightarrow \infty$ with $\Omega_s$ finite we observe a sharp phase transition in the entropy and correspondingly the bipartite entanglement in the system at $\Omega_s = 1$.  
The phase transition characteristic can be seen in an accentuated manner via the plot of derivative of the von Neumann Entropy of the projected bipartite system in Fig.~\ref{Fig:DerivativeEntropyTwoOf}. From the plot it can be noted that the nature of variation of the derivative of the Entropy through a phase transition is similar to the $\lambda$-type variation of the specific heat  observed in second order phase transitions.

We next discuss the role of the phase of the driving field. In fact, we demonstrate that coherent nature of the drive is significant in leading to phase transition behavior. In this case the non-equilibrium steady state density matrix averaged over the random phase possesses only the diagonal elements that are non-zero. The averaged von Neumann entropy of the projected system is plotted in Fig.~\ref{Fig:AveragedEntropyTwoOf}. It can be observed that the transition, even in the asymptotic limit, is not as sharp as in the case of coherent drive and the entropy increases more-or-less in a smooth manner.

From the study of bipartite entanglement it can also be observed that an average two-particle entanglement in the asymptotic limit for large $\Omega_s$  is very close in value to  the bipartite entanglement in the two-particle system giving a clear indication that the two particle entanglement is not a sufficient measure of entanglement in the $N$ particle system. It is furthermore important to note that the popular measure of bipartite entanglement, so-called, concurrence~\cite{Hill:1997, Wootters:1998}
is not a suitable measure for the study of phase transitions~\cite{DickePT}; our conclusions are similar and we therefore do not discuss it here.

It is, of course, important to study the multipartite entanglement in the system as it undergoes NEQPT.  The techniques we  have presented here can be employed to study entanglement of any part, containing $M$ particles, of the total $N$-partite system. We present the results of this study below in the form of the tripartite entanglement.  We note in this case, however, that it is difficult to determine the the asymptotic behavior. 

\begin{figure}
\centerline{\includegraphics[width=0.6\columnwidth]{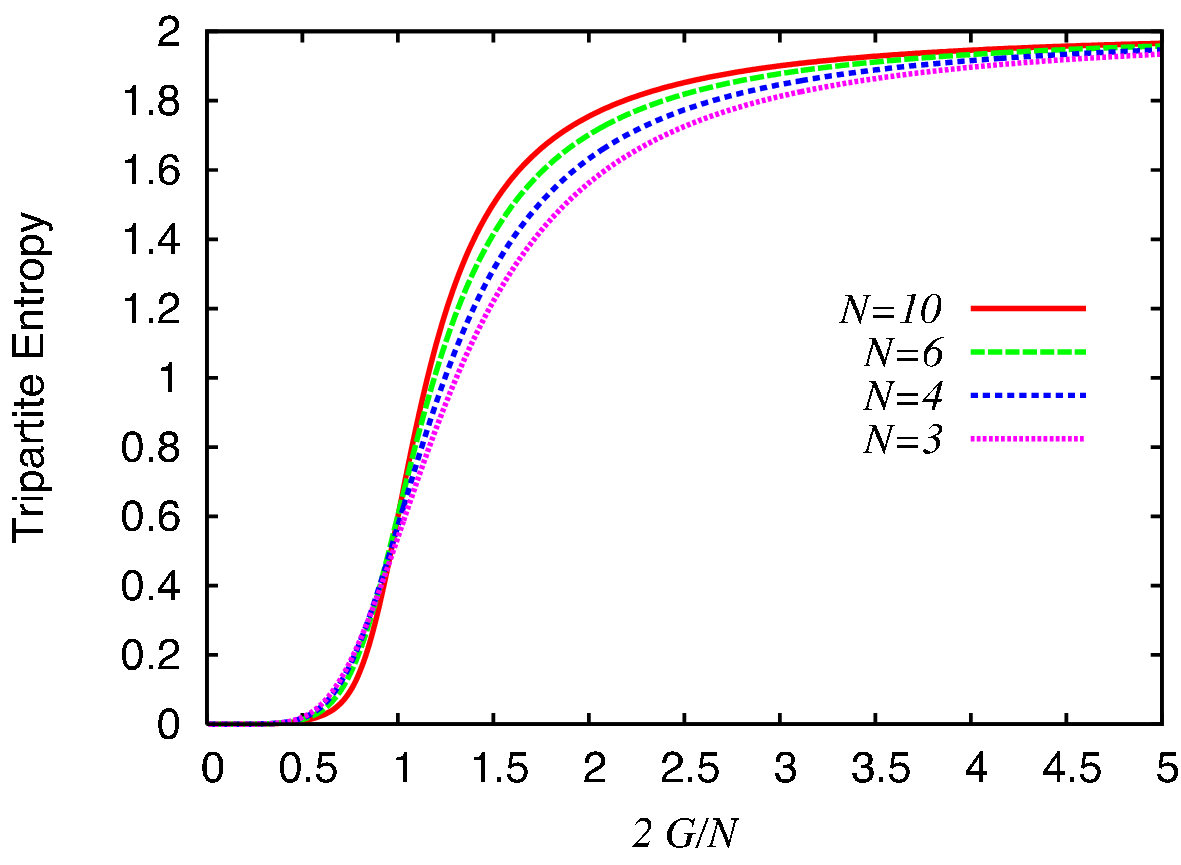}}
\caption{\label{Fig:EntropyThreeOf} Tripartite entanglement in the non-equilibrium QPT examined via the von Neumann Entropy $S(\rho^{(3)})=-\mbox{Tr} \rho^{(3)} \log_2 \rho^{(3)}$.}
\end{figure}

In Fig.~\ref{Fig:EntropyThreeOf} we study tripartite entanglement via the von Neumann Entropy $S(\rho^{(3)})=-\mbox{Tr} \rho^{(3)} \log_2 \rho^{(3)}$ contained in the NESS of the $N$ particle system. The trend of entropy with the increasing value of the coupling parameter $\Omega_s=2 G/N$ is the same as observed for bipartite entanglement in Fig.~\ref{Fig:EntropyTwoOf} except for the sharper features near the transition point $\Omega_s=1$ and larger value of the overall tripartite entropy.
Thus, the result is similar as in the context of spin chains where the entropy of the multiparticle part increases with size in a logarithmic fashion~\cite{Osterloh:2002}. It can be noted that the asymptotic value of the von Neumann entropy of the $M$ partite system in the limit $\Omega_s\rightarrow \infty$ is given by $\log_2(M+1)$ where the state of the system approaches to a completely mixed state. The asymptotic value of the entropy for bi- and tri- partite systems turns out to be $1.585$ and $2$  respectively and is reconfirmed by the plots.

\begin{figure}
\centerline{\includegraphics[width=0.6\columnwidth]{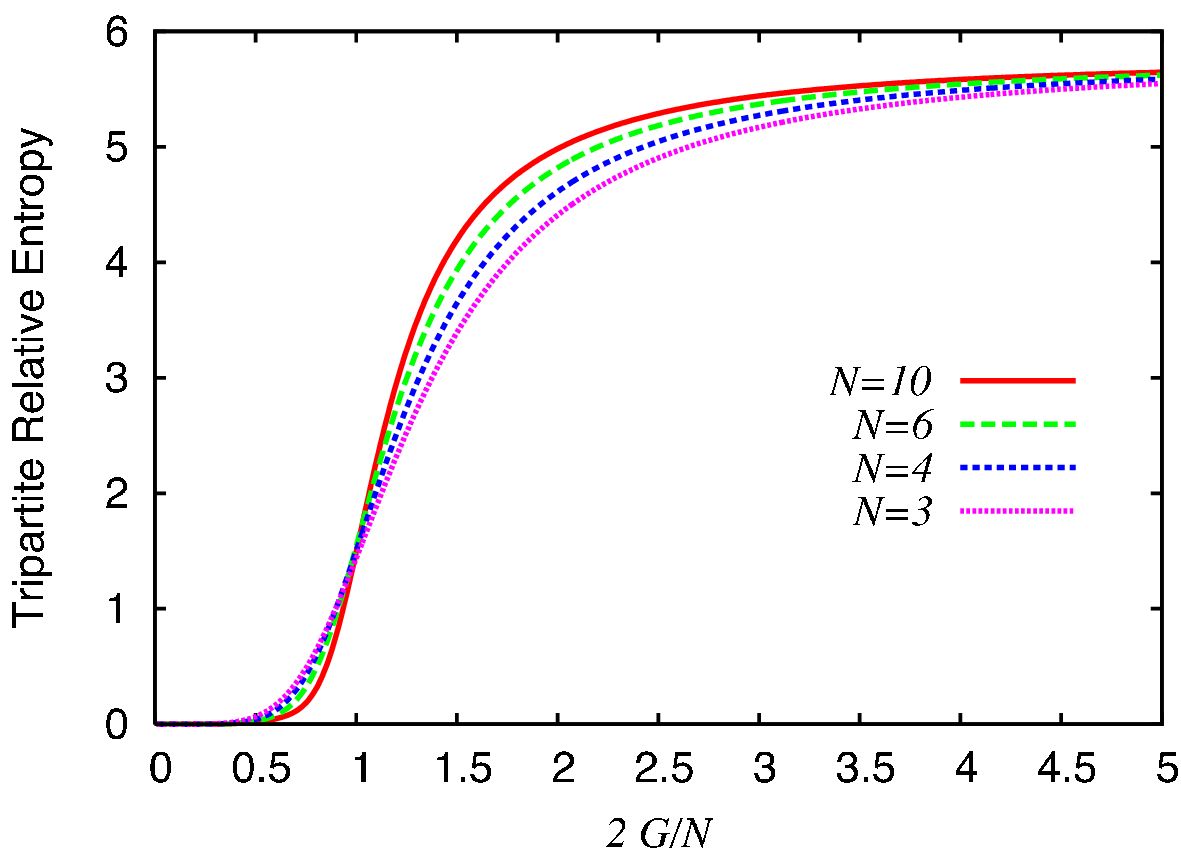}}
\caption{\label{Fig:RelativeEntropyThreeOf} Tripartite entanglement in the non-equilibrium QPT examined via the von Neumann Relative Entropy defined as $3\,S(\rho^{(2)}) - 3\,S(\rho^{(1)})-S(\rho^{(3)})$ where $S$ is defined in the caption of Fig.~\ref{Fig:EntropyThreeOf}}. 
\end{figure}

It is, furthermore, important to study true multiparticle entanglement by separating out entanglement in smaller parts of it. This can be accomplished for example via something called relative entropy.
The relative entropy of a projected $M$ particle density matrix $\rho^{(M)}$ can be evaluated as 
\begin{equation}
S_{R}(\rho^{(M)}) = \sum_{k=1}^M \left[ {M \choose k} S(\rho^{(M-k)})\right]-S(\rho^{(M)})
\end{equation}
where $S(\rho^{(M-k)})$ is the von Neumann Entropy obtained from Eq.~\eqref{Eq:VNE} for the density matrix of $M-k$ particles projected out of the total $N$ particles. The Relative entropy of tripartite entanglement is shown in Fig.~\ref{Fig:RelativeEntropyThreeOf}. The trend is similar to the complete tripartite entanglement, that is with increasing $N$ the transition happens closer to the critical point $\Omega_s=1$ signifying close connection between entanglement and the critical behavior of the NEQPT.

\begin{figure}
\centerline{\includegraphics[width=0.6\columnwidth]{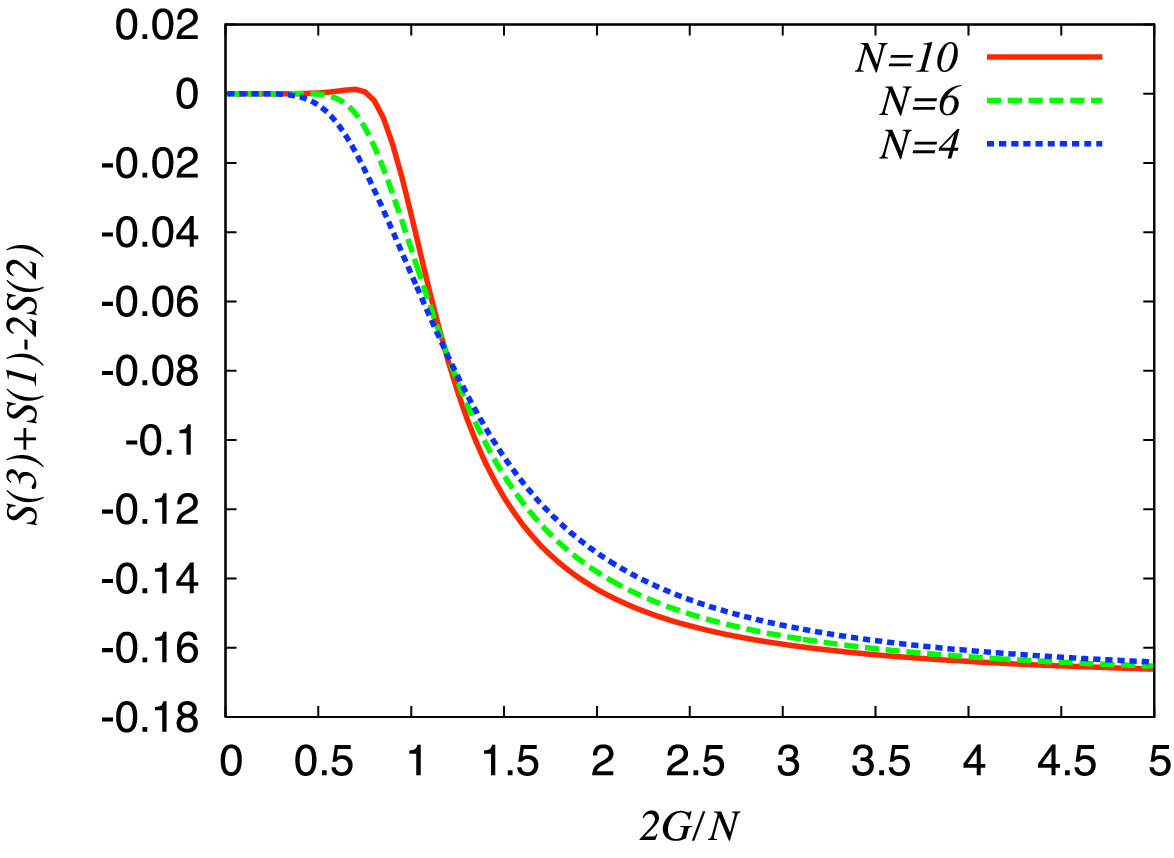}}
\caption{\label{Fig:LiebInequalityEntropy}  Diagnostic of the determined von Neumann Entropy of the parts of the system via the Lieb inequility $S(\rho^{(1)}) + S(\rho^{(3)})<  2 S(\rho^{(2)})$. The negative values of the quantity plotted show that the Lieb inequality is satisfied.}
\end{figure}

As a diagnostic measure of the calculated results we check that the von Neumann entropy of the projected parts of the system satisfies the  Lieb inequality~\cite{Nielsen:2000} in Fig.~\ref{Fig:LiebInequalityEntropy}.
The result is as expected and the quantity $S(\rho^{(1)}) + S(\rho^{(3)})-  2 S(\rho^{(2)})$ remains negative for the whole parameter range.

To summarize, we show realization of a non-equilibrium quantum phase transition and its connection with multipartite entanglement in a collectively driven atomic ensemble that is simultaneously subjected to collective decay. This study has a lot to offer: First of all it allows one to study non-equilibrium quantum phase transitions in simple controllable systems and their associated entanglement properties. It also offers possibility to study correspondence between mesoscopic and macroscopic systems undergoing QPTs. In view of our results it would also be worthwhile to study the first order nonequilibrium phase transitions as in optical bistable systems~\cite{Lugiato:1984} from the point of view of quantum entanglement.

GSA thanks the National Science Foundation, grant no NSF-CCF 0524673, for financial support.


\end{document}